\documentclass[twocolumn, showpacs,amsmath,amssymb,aps,pra,floatfix, 10pt]{revtex4-1}
\usepackage{mathrsfs}
\usepackage{hyperref}
\usepackage{multirow}
\usepackage{bm, epsfig}
\usepackage{graphicx}
\usepackage{subeqnarray}
\usepackage{bbold}
\usepackage{upgreek}
\hypersetup{colorlinks=true, urlcolor=blue, citecolor=blue, filecolor=blue, linkcolor=blue}

\newcommand{\aZ}{\alpha Z}

\begin{document}

\title{Hyperfine splitting in simple ions for the search of the variation of fundamental~constants}

\date{\today}

\author{Natalia~S.~Oreshkina}
\email{oresh@mpi-hd.mpg.de}
\affiliation{Max-Planck-Institut f\"{u}r Kernphysik, Saupfercheckweg 1, 69117 Heidelberg, Germany}
\author{Stefano~M.~Cavaletto}
\affiliation{Max-Planck-Institut f\"{u}r Kernphysik, Saupfercheckweg 1, 69117 Heidelberg, Germany}
\author{Niklas~Michel}
\affiliation{Max-Planck-Institut f\"{u}r Kernphysik, Saupfercheckweg 1, 69117 Heidelberg, Germany}
\author{Zolt\'{a}n~Harman}
\affiliation{Max-Planck-Institut f\"{u}r Kernphysik, Saupfercheckweg 1, 69117 Heidelberg, Germany}
\author{Christoph~H.~Keitel}
\affiliation{Max-Planck-Institut f\"{u}r Kernphysik, Saupfercheckweg 1, 69117 Heidelberg, Germany}

\begin{abstract}
Numerous few-electron atomic systems are considered, which can be effectively used for observing a potential variation of the fine-structure 
  constant $\alpha$ and the electron-proton mass ratio $m_e/m_p$. 
We examine optical magnetic dipole transitions between hyperfine-structure components in heavy highly charged H-like and Li-like ions with observably 
  high sensitivity to a variation of $\alpha$ and $m_e/m_p$.
The experimental spectra of the proposed systems consist of a strong single line, which simplifies significantly the data analysis and 
  shortens the necessary measurement time.
Furthermore, we propose systems for an experimental test of the variation of quark masses and discuss the expected level of accuracy in
  assessing its limitations. 
Finally, we  establish which constraints on the variation of these fundamental constants could be provided
by  measurements with a hyperfine-structure highly-charged-ion clock and some reference clock, showing that a significant improvement of
the current limitations can be reached. 
\end{abstract}


\maketitle



The variation of fundamental constants has been discussed in theories beyond the Standard Model, in unification theories and as a 
  feature of cosmological evolution \cite{Marciano_PRL_1984, Carroll_PRL_1998, Uzan_RMP_2003}. 
Experiments have focused on dimensionless fundamental constants, such as the fine-structure constant $\alpha$, the electron-proton mass ratio 
  $m_e/m_p$, and the light quark masses on the quantum chromodynamic scale $m_q/\Lambda_{\rm QCD}$.
Astrophysical spectra from objects on a distance of a  few billion light years are consistent with predictions based on a variation 
  of the electron-proton mass ratio \cite{Reinhold_PRL_2006} and the fine-structure constant \cite{Webb_PRL_1999, 
  Murphy_MNRAS_2003, Srianand_PRL_2004}, but the assigned errors are on the order of 100\%. 
Latest studies suggest that $\alpha$ is changing not only in time, but also in space \cite{Webb_PRL_2011}.
In addition to astrophysical measurements, also high-precision laboratory measurements of atomic and molecular transitions would allow one to observe 
  or to set limitations on possible variations of the fundamental constants in time and space.

The sensitivity to the electron-proton mass ratio can only assume three given values depending on the type of the considered molecular or atomic transition  
  \cite{Schiller_PRA_2005, Ludlow_RMP_2015}. 
In the case of the variation of the quark masses, nuclear-structure-sensitive transitions need to be considered.
The sensitivity to the variation of the fine-structure constant strongly depends on the chosen system and   can vary by several orders of magnitude.
Consequently, this constant has received a great deal of interest from both experimental and theoretical sides  \cite{Andreev_PRL_2005, 
  Schiller_PRL_2007, Berengut_PRL_2010, Safronova_PRL_2014}.
Probing the variation of $\alpha$ in highly charged ions (HCIs) has two great advantages. 
First, since electrons in HCIs are relativistic, they are very sensitive to the $\alpha$-variation; second, being tightly bound, they are also 
  much less sensitive to external perturbations than neutral atoms or low-charged ions, leading to smaller possible systematic effects. 
Moreover, HCIs already proved to be important and auspicious systems for high-precision tests of quantum electrodynamics (QED) 
  \cite{Shabaev2006109, Shabaev_JPhysCh_2015, Volotka_PRL_2012}, in laboratory astrophysical measurements \cite{Schnorr2013, 
  Oreshkina_prl_2014, Oreshkina_jpb_2016, Simon2010}, and in the precise determination of fundamental constants \cite{Sturm_nature_2014,
  PhysRevLett.96.253002, PhysRevLett.116.100801, Volotka_PRL_2014, Heisse_arxiv_2017}. 
Therefore, outstanding experimental techniques are already developed in HCI research.

Recently, activities towards the experimental observation of a potential $\alpha$-variation  have been started 
  \cite{Windberger_PRL_2015, Bekker2015} on {\it inter}configuration transitions of highly charged Ir$^{17+}$~\cite{Berengut_PRL_2011}.
However, its extremely rich spectrum renders the lines identification very challenging; from the theoretical side, the large amount of interacting electrons 
 limits severely the accuracy of predictions.
Other possible ions \cite{Berengut_PRL_2010, Safronova_PRL_2014, Dzuba_PRA_2015}, proposed for their high sensitivity to the 
  $\alpha$-variation, have a similarly complex atomic structure. 

The above-mentioned complex systems were determined by exploiting level crossing \cite{Berengut_PRA_2012} and selected as candidates for the experimental observation of 
  $\alpha$-variation due to the fact that they  fulfill the following three conditions \cite{Safronova_PRL_2014, Dzuba_PRA_2015}:
First, they ensure that the transition-energy variation $\Delta E$ caused by the anticipated variation of fundamental constants is larger than the 
  experimental accuracy   with which the transition energy can be determined.
Second, the linewidth $\Gamma$ for the candidate ions is smaller than the transition-energy variation, which leads to  ultranarrow lines. 
This condition, however, has not been considered to be necessary by experimentalists for many years.
Finally, these ions feature optical transitions, as needed for high precision measurements on a $\sim 10^{-19}$ fractional uncertainty level 
  \cite{Safronova_PRL_2014,Dzuba_PRA_2015}.
Despite  the existing proofs \cite{Schiller_PRL_2007, Yudin_PRL_2014} that {\it intra}configuration transitions in simpler systems can also be used as 
  $\alpha$-sensitive atomic clocks, only the  above-mentioned {\it inter}configuration transitions in many-electron ions 
  have received considerable interest from the experimental side. 

In the present Rapid Communication, we demonstrate that {\it intra}configuration transitions are more sensitive to the $\alpha$-variation than previously predicted, and therefore can be 
  efficiently used in laboratory.
We investigate further  transitions between the hyperfine structure (HFS) components of heavy H-like and Li-like HCIs with the   simplest atomic structure  
  \cite{Schiller_PRL_2007, Yudin_PRL_2014}.
Thereby, we enrich the actual list of the systems with observable $\alpha$-sensitivity in order to revive experimental interest in these systems.
Additionally, we show how the variation in time of other important quantities, i.e., the electron-proton mass ratio $m_e/m_p$ and 
  the light quark masses on the quantum chromodynamic scale $m_{q,s}/\Lambda_{\rm QCD}$, contribute to the HFS and 
  can be potentially measured in an experiment.


{\it Hyperfine splitting in heavy H-like and Li-like ions.}
The magnetic-dipole HFS appears as a result of the interaction of electrons with the magnetic moment of the nucleus with spin $I$.
Hence, an energy level with fixed electron angular momentum $j$ splits into sublevels with a total angular momentum 
  $F = I+j, I+j-1, \dots, |I-j|$.
For the ground states of H-like and Li-like ions with $j=1/2$, there are only two sublevels, with total angular momenta $F = I+1/2$ and 
  $F = I-1/2$.
For the ions with nuclear charge $Z$, nuclear spin $I$, and nuclear magnetic moment $\mu/\mu_N$, and the valence electron with principal 
  quantum number $n$ and orbital quantum number $l$, the energy of the HFS of the H-like and Li-like ions can be described with the 
  formula \cite{Shabaev_AtPhys_1999, Korzinin_PhysScr_2005}
\begin{align}\label{e-mu}
E_\mu &= \frac{\alpha (\aZ)^3}{n^3(2l+1)j(j+1)} \frac{m_e}{m_p} \frac{\mu}{\mu_N}\frac{2I+1} {2I} m_ec^2  \notag \\
 &\times \biggl [ A(\aZ)(1-\delta)(1-\varepsilon)  + \frac{1}{Z}B(\aZ) + x_{\rm rad} \biggr ].
\end{align}
Here, $A(\alpha Z)$ is a one-electron relativistic factor, which can be calculated analytically for a point-like nucleus 
  \cite{Shabaev_AtPhys_1999, Korzinin_PhysScr_2005}, or numerically for different finite-size nuclear models with the direct inclusion 
  of the finite-nuclear-size (FNS) correction $\delta$ as $A(\aZ)(1-\delta)$;
  $\varepsilon$ is the nuclear-magnetization distribution correction, also called the Bohr-Weisskopf (BW) correction;
  $x_{\rm rad}$ are QED corrections ($x_{\rm rad,Scr}$ in the case of Li-like ions are screened QED corrections, calculated 
  with effective screening potentials \cite{Oreshkina_PLA_2008}), and $B(\aZ)$ stands for interelectronic-interaction corrections 
  of first order in $1/Z$ (only present for Li-like ions).


{\it Sensitivity to $\alpha$-variation and to $m_e/m_p$-variation.}
To characterize how sensitive a transition energy $E$ is to the potential change of different fundamental constants, the enhancement 
  factors~$K$ are introduced following Refs.~\cite{Ludlow_RMP_2015, Flambaum_PRD_2004, Andreev_PRL_2005, Schiller_PRL_2007,Safronova_PRL_2014} 
  as
\begin{align}\label{eq:e_variation}
\frac{\Delta E}{E} & = K_\alpha \frac{\Delta \alpha}{\alpha} + K_{e/p} \frac{\Delta (m_e/m_p)}{m_e/m_p} \notag \\
  &+ K_q \frac{\Delta (m_q/\Lambda_{\rm QCD})}{m_q/\Lambda_{\rm QCD}} + K_s \frac{\Delta (m_s/\Lambda_{\rm QCD})}{m_s/\Lambda_{\rm QCD}}.
\end{align}

In the point-like-nucleus, H-like approximation, taking into account the analytical expression for $A(\aZ)$ 
\cite{Shabaev_AtPhys_1999, Korzinin_PhysScr_2005}, the enhancement factor $K_\alpha$ for the $1s$ and $2s$ states can be presented in a 
simple manner by a Taylor expansion:
\begin{align}
& K_\alpha \left[\begin{matrix} {\text 1s} \\ {\text 2s} \end{matrix}\right] \equiv \frac{d E}{d \alpha} \bigg/ \frac{E}{\alpha} \notag \\
&= 4 + \left[\begin{matrix} {3} \\ {17/4} \end{matrix}\right] (\alpha Z)^2 + \left[\begin{matrix} {4} \\ {5} \end{matrix}\right]
(\alpha Z)^4 + O\left((\alpha Z)^6\right).
\end{align}
It clearly indicates that HFS transitions for the states considered (and, in fact, for an arbitrary state) have a high sensitivity to 
  a potential $\alpha$-variation, and that the calculated enhancement factor $K_\alpha$   was underestimated in Ref.~\cite{Schiller_PRL_2007}. 
For very high $Z$, FNS and BW effects become important, slightly reducing the value of $K_\alpha$, whereas QED and interelectronic-interaction 
  contributions to $K_\alpha$ were estimated to be negligible.
Additionally, as it is clear from Eq.~\eqref{e-mu}, the sensitivity of the HFS to the potential variation of the electron-proton mass ratio 
$m_e/m_p$ is the same for all the ions and equal to $K_{e/p}=1$.

{\it Sensitivity to the variation of quark masses $m_q/\Lambda_{\rm QCD}$ and $m_s/\Lambda_{\rm QCD}$.}
The procedure to estimate this effect was described in \cite{Flambaum_PRD_2004}.
In the single-particle approximation, the magnetic moment of the nucleus is
\begin{equation}\label{eq:mu_singe}
\frac{\mu}{\mu_N} = \left\{ 
\begin{array}{ll} [g_S + (2I-1)g_L]/2 & \text{for } I=L+1/2,  \\ 
	\cfrac{I}{2(I+1)}[-g_S + (2I+3)g_L] & \text{for } I=L-1/2. \end{array} \right.
\end{equation}
Here, $L$ stands for the orbital angular momentum of the nuclei, while the orbital $g$ factors are $g_L=1$ for the proton and $g_L=0$ for 
  the neutron, and the spin $g$ factors $g_S=5.586$ for the proton and $g_S=-3.826$ for the neutron.
Denoting  $m_q = (m_u + m_d)/2$ the averaged mass of the up and down quarks, and $m_s$ the strange quark mass, one can calculate 
  the QCD corrections to the proton and neutron magnetic momenta  \cite{Flambaum_PRD_2004}:
\begin{align}\label{eq:mu_quark}
\frac{\delta \mu_p}{\mu_p} &= -0.087 \frac{\delta m_q}{m_q}, \notag \\ 
\frac{\delta \mu_p}{\mu_p} &= -0.013 \frac{\delta m_s}{m_s}, \notag \\ 
\frac{\delta \mu_n}{\mu_n} &= -0.118 \frac{\delta m_q}{m_q}, \notag \\ 
\frac{\delta \mu_n}{\mu_n} &= 0.0013 \frac{\delta m_s}{m_s}. 
\end{align}
Therefore, the variation of the quark masses leads to a variation of the nuclear magnetic moment $\mu$, and their potential variations  
  can be also seen in high-precision experiments on the HFS of HCIs.
Using Eqs. \eqref{eq:mu_singe}, \eqref{eq:mu_quark},  we calculated the enhancement factors  $K_q$ and $K_s$    for several ions.

\begin{table*}
\begin{center}
\begin{tabular}{l r@{}l c c c c c r@{}l r@{}l c c c c}
\hline\hline
Ion	& $\mu$&$/\mu_N$ & $I$	& $A(\aZ)$ & $\delta$ & $\varepsilon$ & $x_{\rm rad}$  &$E_\mu$ &  & $\Gamma/$&$(2\pi)$  & $K_\alpha$ 
& {$K_{e/p}$} & {$K_q$} & {$K_s$ }\\ 
 & & & & & & & &  (TH&z) &  (H&z) & \\
\hline
${}^{113}$In${}^{48+}$	& 5.&5289(2)	& 9/2	& 1.2340 & 0.0171 & 0.0048 & -0.0042	
  & 221&.2(3)	& 2.&84  & 4.43	 & 1 & -0.036 & -0.005 \\	
${}^{121}$Sb${}^{50+}$	& 3.&3634(3)	& 5/2	& 1.2581 & 0.0191 & 0.0053 & -0.0047 	
  & 166&.6(3)	& 1.&11  & 4.47	 & 1 & -0.051 & -0.008 \\	
${}^{133}$Cs${}^{54+}$	& 2.&582025(3)	& 7/2	& 1.3125 & 0.0236 & 0.0018 & -0.0050	
  & 159&.2(1)	& 1.&00  & 4.57	 & 1 & 0.110 & 0.016 \\	
${}^{139}$La${}^{56+}$	& 2.&7830455(9)	& 7/2	& 1.3430 & 0.0263 & 0.0026 & -0.0052	
  & 194&.7(2)	& 1.&81  & 4.61	 & 1 & 0.110 & 0.016 \\	
${}^{153}$Eu${}^{62+}$	& 1.&5324(3)	& 5/2	& 1.4509 & 0.0369 & 0.0049 & -0.0061	
  & 162&.0(2)	& 0.&967 & 4.78	 & 1 & -0.051 & -0.008 \\	
${}^{159}$Tb${}^{64+}$	& 2.&014(4)	& 3/2	& 1.4933 & 0.0405 & 0.0070 & -0.0064	
  & 265&.7(8)	& 3.&79  & 4.85	 & 1 & 1.165 & 0.174 \\	
${}^{171}$Yb${}^{69+}$	& 0.&49367(1)	& 1/2	& 1.6170 & 0.0540 & 0.0300 & -0.0073	
  & 127&(1)	& 0.&270 & 5.02	 & 1 & -0.118 & 0.001 \\	
${}^{189}$Os${}^{75+}$	& 0.&659933(4)	& 3/2	& 1.8092 & 0.0749 & 0.0216 & -0.0085	
  & 160&(1)	& 0.&773 & 5.26	 & 1 & -0.118 & 0.001 \\	
${}^{195}$Pt${}^{77+}$	& 0.&60952(6)	& 1/2	& 1.8874 & 0.0836 & 0.0377 & -0.0089	
  & 243&(3)	& 1.&79	 & 5.36	 & 1 & -0.118 & 0.001 \\	
${}^{197}$Au${}^{78+}$	& 0.&1469(13)$^a$ & 3/2	& 1.9297 & 0.0884 & 0.0115 & -0.0091	
  & 42&.5(6)	& 0.&0142 & 5.40 & 1 & 1.165 & 0.174 \\	
${}^{199}$Hg${}^{79+}$	& 0.&5058855(9)	& 1/2	& 1.9744 & 0.0934 & 0.0402 & -0.0094	
  & 225&(3)	& 1.&39	 & 5.45	 & 1 & -0.118 & 0.001 \\	
${}^{207}$Pb${}^{81+}$	& 0.&592583(9)	& 1/2	& 2.0718 & 0.1047 & 0.0426 & -0.0098	
  & 294&(4)	& 3.&04  & 5.56	 & 1 & -0.118 & 0.001 \\	
${}^{235}$U${}^{91+}$	& 0.&38(3)	& 7/2	& 2.7978 & 0.1899 & 0.0106 & -0.0125	
  & 187&(15)	& 1.&24	 & 6.16	 & 1 & -0.118 & 0.001 \\	
\hline\hline
\end{tabular}
\caption{Individual contributions to the hyperfine splitting (see Eq.~\eqref{e-mu}), the transition energy $E_\mu$, 
  the linewidth $\Gamma/(2\pi)$ of the excited state, and the sensitivity factors $K_\alpha$, $K_{e/p}$,   $K_q$, and $K_s$  
  for the ground state of H-like ions.\label{tab:HFS_Hlike}}
\end{center}
\end{table*}

\begin{table*}
\begin{tabular}{l r@{}l c c c c r@{}l r@{}l c c c c}
\hline\hline
Ion	& $\mu$&$/\mu_N$ & $I$ & $a(\aZ)$ & $B(\aZ)$ & $x_{\rm rad, Scr}$ &  $E_\mu$ & &  $\Gamma/$&$(2\pi)$ & $K_\alpha$ 
& {$K_{e/p}$} & {$K_q$} & {$K_s$} \\	
 & & &&  & & &  (TH&z) &  (H&z) & \\
\hline
${}^{175}$Lu${}^{68+}$	& 2.&2323(11)	& 7/2	& 1.8740 & -0.0739 & -0.0085	
  & 51&.79(3)	& 0.&0369 & 5.47 & 1 & 0.110 & 0.016 \\	
${}^{185}$Re${}^{72+}$	& 3.&1871(3)	& 5/2	& 1.9963 & -0.0767 & -0.0098	
  & 98&.5(4)	& 0.&240 & 5.69	 & 1 & -0.051 & -0.008 \\	
${}^{187}$Re${}^{72+}$	& 3.&2197(3)	& 5/2	& 1.9960 & -0.0767 & -0.0098	
  & 99&.5(4)	& 0.&247 & 5.69	 & 1 & -0.051 & -0.008 \\	
${}^{203}$Tl${}^{78+}$	& 1.&622257	& 1/2	& 2.2820 & -0.0826 & -0.0119	
  & 120&.6(7)	& 0.&261 & 6.07	 & 1 & -0.087 & -0.013 \\	
${}^{209}$Bi${}^{80+}$	& 4.&1106(2)	& 9/2	& 2.4065 & -0.0852 & -0.0127	
  & 193&(1)	& 1.&91  & 6.22	 & 1 & 0.076 & 0.011 \\	
${}^{231}$Pa${}^{88+}$	& 2.&01(2)	& 3/2	& 2.9783 & -0.0990 & -0.0167	
  & 185&(2)	& 1.&38  & 6.86	 & 1 & -0.064 & -0.010 \\	
\hline\hline
\end{tabular}
\caption{Individual contributions to the hyperfine splitting (see Eq.~\eqref{e-mu}, $a(\aZ) = A(\aZ)(1-\delta)(1-\varepsilon)$), 
  the transition energy $E_\mu$, the linewidth $\Gamma/(2\pi)$ of the excited state, and sensitivity factors $K_\alpha$, $K_{e/p}$,   $K_q$, 
  and $K_s$  for the ground state of Li-like ions. For ${}^{209}$Bi${}^{80+}$ ion, the BW correction was calculated using the 
  specific-difference approach \cite{Shabaev_AtPhys_1999} with experimental data from~\cite{Klaft_PRL_1994}. \label{tab:HFS_Lilike}}
\end{table*}


{\it Theoretical predictions for HFS parameters.}
The individual contributions to the energy of H-like HFS, the HFS energy, the linewidth $\Gamma/(2\pi)$ of the excited state, 
  calculated with the formulas from Ref.~\cite{Shabaev_AtPhys_1999}, and the enhancement factors $K_{e/p}$, 
  $K_q$, $K_s$, and $K_\alpha$, calculated with the inclusion of FNS and BW effects for several H-like ions, are listed in 
  Table \ref{tab:HFS_Hlike}, and for Li-like ions in Table \ref{tab:HFS_Lilike}. 
The nuclear parameters were taken from \cite{Angeli_2013,Stone_2005}.
The accuracy of the theoretical predictions of HFS is mostly limited by our knowledge of the nuclear magnetic moment and by the uncertainty 
  of the BW contribution $\varepsilon$ \cite{Shabaev_AtPhys_1999}.
Higher-order contributions, such as second-order QED corrections, two-photon exchange and rigorous screened QED corrections for Li-like ions, 
  were found earlier~\cite{Volotka_PRL_2012} to be under the BW-correction uncertainty level, so they can be ignored here.
The HFS spectra of these systems consist of strong single optical line and feature small systematic effects.
Therefore, despite the relatively high uncertainty in the theoretical predictions, they can be and have been measured experimentally with a high accuracy 
  \cite{Beiersdorfer_PRL_1998,Lochmann_PRA_2014, NatCom_Bi_2017}.

The enhancement factor $K_{e/p}$ is equal to 1 for the HFS transitions for all the ions considered, the enhancement factors $K_q$ and $K_s$ 
  are on the level of 0.1 and less  for all ions, except for ${}^{159}$Tb${}^{64+}$ and ${}^{197}$Au${}^{78+}$ with $K_q=1.165$.
Therefore, the most relevant question is how good the HFS of HCIs is for the search of a potential $\alpha$ variation in comparison with 
  other atomic systems.
As one can see from Table \ref{tab:HFS_Hlike} and Table \ref{tab:HFS_Lilike}, the $\alpha$-sensitivity coefficient $K_\alpha$ 
  is about 5-6 for all the ions considered here. 
In comparison, by exploiting the level crossing one could find ions featuring narrow optical transitions with $K_\alpha$ up to 
  10-100~\cite{Berengut_PRL_2011, Safronova_PRL_2014, Dzuba_PRA_2015}.
However, for each isoelectronic sequence such transitions can be found only for one, rarely  two HCIs.   
This extremely restricts  the choice of atomic systems available for experimental observations. 
Conversely, for H-like and Li-like HFS there are no such limitations, and any of the ions in Tables~\ref{tab:HFS_Hlike} and 
  \ref{tab:HFS_Lilike} (and other HCIs not listed there) can be used.
Therefore, it is possible to choose the most suitable ion with respect to experimentally important parameters.

In the ground states of H-like and Li-like systems, the first-order quadrupole shift, one of the most important systematic uncertainties, vanishes completely, since the 	total angular momentum of the electrons is equal to $1/2$ (see also Supplemental Matherials).
Other systematic effects (magnetic field, black body radiation, Doppler effects, gravity) for HCIs have been discussed in 
  \cite{Schiller_PRL_2007, Derevianko_PRL_2012, Yudin_PRL_2014}   and shown to contribute less than $10^{-19}-10^{-20}$ to the 
  fractional uncertainty of the transition energy. 
The  transition energies for most of the ions in Tables \ref{tab:HFS_Hlike} and \ref{tab:HFS_Lilike} are in the range of high-stability 
  lasers, while the natural linewidth of $\sim$1~Hz is lower than the bandwidth of an ultra-stable laser.
This is not a disadvantage compared to the ultranarrow lines considered in Refs.~\cite{Berengut_PRL_2011, Safronova_PRL_2014, Dzuba_PRA_2015}, since the resolution of 
  the observed lines is mostly determined by the bandwidth of the laser, as we show in the Supplemental Materials.
Furthermore, the excitation of broader lines requires much lower laser intensities with a strong suppression of AC-Stark shifts.
With the anticipated  potential variation of $\Delta\alpha/\alpha$ on the level $10^{-18}-10^{-19}$/year, and the enhancement factor of 
  $K_\alpha \approx 6$, the relative change in the transition energy HFS HCI will be on a 
  level of $6-60\times 10^{-19}$/year, which is 6 to 60 times larger than the predicted experimental fractional accuracy of the transition 
  energy \cite{Schiller_PRL_2007,Yudin_PRL_2014}.


{\it Variation of the quark masses.}
In order to illustrate how HFS of HCIs can be used to investigate the variation of the quark masses,  we 
  consider the HFS clock ${}^{159}$Tb${}^{64+}$ and the reference HFS clock ${}^{153}$Eu${}^{62+}$,
  sensitive and insensitive, respectively, to the variation of the quark masses.
$K_{e/p}=1$ for both clocks and their $\alpha$-sensitivities are very close, leading to a low relative $\alpha$-sensitivity 
  $K_{\alpha}[{\rm Tb}] - K_{\alpha}[{\rm Eu}] = 0.07$, whereas the
  relative sensitivity to the variation of the quark masses is large, $K_{q}[{\rm Tb}] - K_{q}[{\rm Eu}] = 1.216$.
A combination of those two clocks would hence provide a unique opportunity to test the variation of the quark masses  
  via atomic transitions in the optical domain. 
Also another pair of ions, ${}^{197}$Au${}^{78+}$ and ${}^{199}$Hg${}^{79+}$, would provide the same experimental test.
The comparison of two HFS clocks, sensitive  and insensitive to the variation of the quark masses on the level of $10^{-17}/$yr, would immediately provide
  the same uncertainty level in assessing limitations on the variation of the quark masses $\frac{m_q}{\Lambda_{\rm QCD}}$, 
  since the relative sensitivity coefficient is $K_{q} \approx 1$.

{\it Variation of the fine-structure constant and electron-proton mass ratio.} 
Once limits on the variation of quark masses are discussed, one can focus on the experimental search for the variation of 
  the fine-structure constant $\alpha$ and the electron-proton mass ratio $m_e/m_p$.
By combining one of our proposed  HFS of HCIs clocks (for  simplicity let us consider the clock with a small sensitivity $K_{q,s}$  to the variation of quark masses) 
  with other existing clocks, one could  explore the possible variation of $\alpha$ and $m_e/m_p$  \cite{Ludlow_RMP_2015}. 
The relative variation of the HFS energies is always linear in $\frac{\delta(m_e/m_p)}{m_e/m_p}$, see Eq.~\eqref{eq:e_variation}, for 
  any HFS-based system.
Therefore, in order to isolate the dependence of HFS on $\alpha$-variation, the HCI clocks we put forward should be compared with another 
  clock system also based on HFS. 
This would allow one to cancel the contribution of the $m_e/m_p$ variation, enabling one to see pure $\alpha$-variation in time.
By comparing it with clocks not based on HFS, e.g., the electric octupole transition in Yb with  
  $K_\alpha = -5.95$ \cite{Ludlow_RMP_2015}, one can have a system with an enhanced effective  sensitivity of~\footnote{the term $-2$ originates from the 
  different definitions of the sensitivity coefficient $K_\alpha$ in different references - with or without the separation of the Rydberg constant in the transition energy}
  $K_\alpha^{\rm HFS} - K_\alpha^{\rm Yb} -2 \approx 10$, 
  and provide additional constraints for $\frac{\delta \alpha}{\alpha}$ and $\frac{\delta(m_e/m_p)}{m_e/m_p}$ \cite{Huntemann_PRL_2014}.
Moreover, taking the linear combination of the transition energies of clocks based on HFS and clocks not based on HFS, one may have  a 
  constraint on $\frac{\delta(m_e/m_p)}{m_e/m_p}$ only.
Though the sensitivity of the proposed HFS clocks is of the same order as the sensitivity of Yb E3 clocks, the signs of those two contributions 
  are opposite. 
Therefore, both clocks combined together would lead to an additional enhancement factor of around~2.

The current limitations \cite{Ludlow_RMP_2015} for the variations are $\frac{\delta \alpha}{\alpha} = -2.0(2.0)\times 10^{-17}/$yr and 
  $\frac{\delta(m_e/m_p)}{m_e/m_p} = 0.5(1.6)\times 10^{-16}/$yr, the number in the parenthesis gives the uncertainty. 
The possible opposite signs of these variations mean that, in a HFS-based clock, observing the variation of the fine-structure constant can be hindered by the 
  variation of the electron-proton mass ratio, since, in this case, they partly cancel each other. 
However, the relatively large error bars for both variations allow the contributions to add as well, leading to an enhanced sensitivity. 
Moreover, different linear combinations, mentioned before, could provide new, more accurate limitations for the variation of both 
  fundamental constants.  

\begin{figure}
\centering
\includegraphics[width=0.99\columnwidth, keepaspectratio]{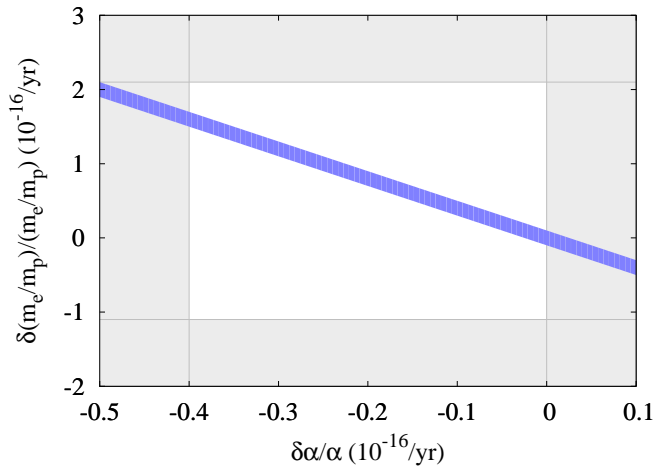}
\caption{(Color online) The current limits~\cite{Ludlow_RMP_2015} for the variation of the fine-structure constant 
  $\frac{\delta \alpha}{\alpha} = -2.0(2.0)\times 10^{-17}/$yr and electron-proton mass ratio 
  $\frac{\delta(m_e/m_p)}{m_e/m_p} = 0.5(1.6)\times 10^{-16}/$yr (white rectangle) together with the new limitations based on the comparison of HFS clocks
  with insensitive atomic clocks on the level of $10^{-17}/$yr, given by Eq.~\eqref{eq:limits} (blue band).
}
\label{fig:limits}
\end{figure}

{\it New limits on the variation of the fundamental constants with the HFS of HCIs.}
Finally, let us establish which constraints on the variation of the fundamental constants could be provided by the measurements with a HFS 
  HCI clock and some reference clock.
In a conservative scenario, we choose insensitive ($K_\alpha^{\rm insen.}=K_{e/p}^{\rm insen.}=0$) atomic clocks.   
The accuracy level of both HFS and reference clocks are also conservatively assumed to be only $10^{-17}/$yr, 2 or 3 orders of
  magnitude lower than the estimated systematic effects for the HFS of HCIs \cite{Schiller_PRL_2007, Yudin_PRL_2014}.
Then the comparison of the proposed HFS clock ($K_\alpha^{\rm HFS}=6$, $K_{e/p}^{\rm HFS}=1$, $K_{q,s}^{\rm HFS}\approx 0$) with the reference atomic clock 
  on the chosen level of accuracy would lead to a new limitation:
\begin{equation}\label{eq:limits}
  \left| (K_\alpha^{\rm HFS} - K_\alpha^{\rm insen.} -2)\frac{\delta \alpha}{\alpha} + \frac{\delta(m_e/m_p)}{m_e/m_p} \right| < 
  10^{-17}/\text{yr}.
\end{equation}
The current limit for the variation of the fine-structure constant $\alpha$ and the electron-proton mass ratio ${m_e/m_p}$ together with 
  new possible constraints are plotted in Fig.\ref{fig:limits}. 
The improvement of the current limitations can be clearly seen even for the assumed conservative scenario.
Any advantage either in the enhanced sensitivity of two clocks or in their accuracy would provide even more stringent limitations 
  on the possible variation of fundamental constants.


%
{\it Conclusions and outlook.}
We put forward  intraconfiguration transitions that can be used in laboratory search for a potential variation of $\alpha$, $m_e/m_p$, and even 
  $m_q/\Lambda_{\rm QCD}$.
Laboratory measurements of the HFS of simplest atomic systems, namely, of H-like and Li-like ions, with currently available  experimental 
  techniques, can be performed to the 19th decimal, allowing one to observe those potential variations.
The transition-energy variation caused by an  anticipated change of the fine-structure constant in the time interval of one year 
  is estimated to be 6 to 60 times larger than the expected experimental accuracy of the transition energy.
Simple strong-single-line HFS spectra in HCI advantageously simplify the experimental observation and identification of the lines.
Additionally, the use of the considered hyperfine $M1$-type splitting instead of ultranarrow transitions in configuration-level-crossing 
  systems would allow one to reduce the measurement time due to the absence of a quadrupole splitting and the ability to use a much  
  weaker laser for the excitation of the ground state, suppressing AC-Stark shifts.


{\it Acknowledgements.}
The authors acknowledge helpful discussions with D. A. Glazov, J. R. Crespo L\'opez-Urrutia, L.~Schm\"{o}ger, and V.~Debierre.

\bibliography{refs}


\end{document}